\documentstyle[12pt]{article}
\makeatletter

\if@twoside
Probably too small to be useful
\else
\fi
\ifcase \@ptsize
\or 
\or 
\fi

\def\section{\@startsection {section}{1}{\z@}{-3.5ex plus -1ex minus
 -.2ex}{2.3ex plus .2ex}{\bf\raggedright}}
\def\subsection{\@startsection{subsection}{2}{\z@}{-3.25ex plus -1ex minus
 -.2ex}{1.5ex plus .2ex}{\sc\raggedright}}
\def\bibliography#1{\if@filesw\immediate\write\@auxout
  {\string\bibstyle{npb}}\fi
  \if@filesw\immediate\write\@auxout{\string\bibdata{#1}}\fi
  \@input{\jobname.bbl}}
\def\thebibliography#1{\section{References\@mkboth
 {REFERENCES}{REFERENCES}}\list
 {[\arabic{enumi}]}{\settowidth\labelwidth{[#1]}\leftmargin\labelwidth
 \advance\leftmargin\labelsep \itemsep=0pt
 \usecounter{enumi}}
 \def\newblock{\hskip .11em plus .33em minus .07em}
 \sloppy\clubpenalty4000\widowpenalty4000
 \sfcode`\.=1000\relax}
\def\@citex[#1]#2{%
\if@filesw \immediate \write \@auxout {\string \citation {#2}}\fi
\@tempcntb\m@ne \let\@h@ld\relax \def\@citea{}%
\@cite{%
  \@for \@citeb:=#2\do {%
    \@ifundefined {b@\@citeb}%
      {\@h@ld\@citea\@tempcntb\m@ne{\bf ?}%
      \@warning {Citation `\@citeb ' on page \thepage \space undefined}}%
      {\@tempcnta\@tempcntb \advance\@tempcnta\@ne%
      \@tempcntb\number\csname b@\@citeb \endcsname \relax%
      \ifnum\@tempcnta=\@tempcntb 
        \ifx\@h@ld\relax%
          \edef \@h@ld{\@citea\csname b@\@citeb\endcsname}%
        \else%
          \edef\@h@ld{\ifmmode{-}\else--\fi\csname b@\@citeb\endcsname}%
        \fi%
      \else
        \@h@ld\@citea\csname b@\@citeb \endcsname%
        \let\@h@ld\relax%
      \fi}%
    \def\@citea{,\penalty\@highpenalty\,}%
  }\@h@ld%
}{#1}}
\newif\ifabstract \abstractfalse
\newif\ifconsolidatetitle \consolidatetitlefalse
\def\ds@consolidatetitle{\consolidatetitletrue}
\gdef\@publabel{\hfil}
\gdef\@pubdate{\null}
\gdef\@pubnumber{\null}
\gdef\@author{\null}
\gdef\@title{\null}
\gdef\@abstract{\null}
\long\def\pubdate#1{\gdef\@pubdate{#1}}
\long\def\pubnumber#1{\gdef\@pubnumber{#1}}
\long\def\publabel#1{\gdef\@publabel{#1}}
\long\def\author#1{\gdef\@author{#1}}
\long\def\title#1{\gdef\@title{#1}}
\long\def\abstract#1{\abstracttrue\gdef\@abstract{#1}}
\def\titlerelax{
\let\maketitle\relax
\let\settitleparameters\relax
\let\consolidatetitle\relax
\let\inittitlepage\relax
\let\finishtitlepage\relax
\let\titlepagecontents\relax
\let\multithanks\relax
\let\titlebaselines\relax
\let\@makepub\relax
\let\@maketitle\relax
\let\@makeauthor\relax
\let\@makeabstract\relax
\let\@maketitlenote\relax
\let\thanks\relax
\let\titlerelax\relax}
\def\titleclean
{\gdef\@titlenote{}
\gdef\@abstract{}
\gdef\@author{}
\gdef\@title{}
\gdef\@pubdate{}\gdef\@pubnumber{}\gdef\@publabel{}
\gdef\@dpublabel{}
}

\def\@makepub{\vbox to \z@{\hbox to \textwidth{\hfill
\@publabel \hfill
\llap{\parbox[t]{0.25\textwidth}{\raggedleft\@pubnumber}}}%
\vss}}
\def\@maketitle{\vskip 60pt \begin{center}
 {\LARGE \@title \par}
 \end{center}}
\def\@makeauthor{{%
\def\and{\smallskip {\normalsize \rm and\smallskip }}
\def\And{\medskip {\normalsize \rm and\\}\medskip}
\long\def\address##1{{\def\and{\\and\\}\medskip
				{\small \it \\##1\\}
}}
{\centering
 \vskip 3em
 \large \lineskip .75em
 \@author}
 \par}}
\def\@makedate{\vskip 1.5em
 {\raggedright \small \noindent\@pubdate \par}}
\def\@makeabstract{\vskip 1.5em
{\small
\begin{center}
{\bf ABSTRACT\vspace{-.5em}\vspace{0pt}}
\end{center}
\quotation \@abstract \endquotation}}
\def\@consolidatetitle{{
\thispagestyle{empty}
\@makepub
\setlength{\parskip}{0pt}
\null
\vskip 7mm
\nointerlineskip
\@maketitle
\vskip 5ex
\@makeauthor
\vskip 4 ex
\@makeabstract
\vskip 5 ex
}}

\def\maketitle{\ifconsolidatetitle\@consolidatetitle
\else
	\titlepage
	\let\footnotesize\small \setcounter{page}{0}
	\@makepub
	\vfil
	\@maketitle
	\@makeauthor
	\vfil
	\ifabstract\@makeabstract\fi
	\@thanks
	\vfil
	\@makedate
	\if@restonecol\twocolumn\fi \eject
\fi
	\titlerelax \titleclean
	\setcounter{footnote}{0}
}

\def\square{\hbox{\vrule height2ex\kern-0.4pt
\vbox to 2ex{\hrule width2ex\vfil\hrule width2ex}\kern-0.4pt\vrule height2ex}}

{\unskip\nobreak\hfill$\Box$\end{list}}

\def\lefteqn#1{\hbox to 0pt{$\displaystyle #1$\hss}}
\def\starteqn#1#2{\hbox to 0pt{${\displaystyle #1}\textstyle #2$\hss}
\qquad\nonumber\\*&#2&}
\def\b{\beta}

\def\blank#1{}
\def\cdd{{\cdot}}

\def\en{\end{equation}}

\def\enn{\end{eqnarray}}

\def\eq{\begin{equation}}
\def\eqq{\begin{eqnarray}}

\def\half#1{\frac{#1}{2}}
\def\hf{\half 1}

\def\vev#1{\left\langle#1\right\rangle}

\def\eqq{\begin{eqnarray}}
\def\enn{\end{eqnarray}}

\ifcase\@ptsize
 \font\tenmsa=msam10
 \font\sevenmsa=msam7
 \font\fivemsa=msam5
 \font\tenmsb=msbm10
 \font\sevenmsb=msbm7
 \font\fivemsb=msbm5
\or
 \font\tenmsa=msam10 scaled \magstephalf
 \font\sevenmsa=msam8
 \font\fivemsa=msam6
 \font\tenmsb=msbm10 scaled \magstephalf
 \font\sevenmsb=msbm8
 \font\fivemsb=msbm6
\or
 \font\tenmsa=msam10 scaled \magstep1
 \font\sevenmsa=msam8
 \font\fivemsa=msam6
 \font\tenmsb=msbm10 scaled \magstep1
 \font\sevenmsb=msbm8
 \font\fivemsb=msbm6
\fi

\newfam\msafam
\newfam\msbfam
\textfont\msafam=\tenmsa  \scriptfont\msafam=\sevenmsa
  \scriptscriptfont\msafam=\fivemsa
\textfont\msbfam=\tenmsb  \scriptfont\msbfam=\sevenmsb
  \scriptscriptfont\msbfam=\fivemsb

\def\hexnumber@#1{\ifnum#1<10 \number#1\else
 \ifnum#1=10 A\else\ifnum#1=11 B\else\ifnum#1=12 C\else
 \ifnum#1=13 D\else\ifnum#1=14 E\else\ifnum#1=15 F\fi\fi\fi\fi\fi\fi\fi}

\def\msa@{\hexnumber@\msafam}
\def\msb@{\hexnumber@\msbfam}
\def\Bbb{\ifmmode\let\next\Bbb@\else
 \def\next{\errmessage{Use \string\Bbb\space only in math mode}}\fi\next}
\def\Bbb@#1{{\Bbb@@{#1}}}
\def\Bbb@@#1{\fam\msbfam#1}

\makeatother

\begin{document}

\pubnumber{DTP 92--23 \\DAMTP 92--41}
\pubdate{20 June 1992}
\title{$G_2^{(1)}$ Affine Toda Field Theory:\\ A Numerical Test of Exact
S--Matrix Results}
\author{G.~M.~T.~WATTS$^{1,2}$
and
ROBERT~A.~WESTON$^{3,4}$
\address{
Department of Mathematical Sciences, \\
University of Durham, South Road,
Durham, DH1 3LE, U.K.}
}

\footnotetext[2]{
Email: {\tt G.M.T.Watts@UK.AC.DURHAM}
}

\footnotetext[2]{Address from 1st July: DAMTP, University
of Cambridge, Silver Street, Cambridge, U.K.}

\footnotetext[3]{
Email: {\tt R.A.Weston@UK.AC.DURHAM}
}

\footnotetext[4]{Address from 1st Oct:
Centre de Recherches Math\'ematiques, Universit\'e de Montr\'eal,
C.P.~6128--A, Montr\'eal, Qu\'ebec, H3C 3J7, Canada.}

\abstract{
We present the results of a Monte--Carlo simulation of the $G_2^{(1)}$
Affine Toda field theory action in two dimensions. We measured the
ratio of the masses of the two fundamental particles as a function of
the coupling constant. Our results strongly support the conjectured
duality with the $D_4^{(3)}$ theory, and are consistent with the
mass formula of Delius et al.}


\maketitle

 \def\reff#1{(\ref{#1})}

\section{Introduction \label{secv1}}

An affine Toda field theory is a theory of scalar fields in two
dimensions with exponential interactions. There is an affine Toda field
theory associated with each Kac--Moody algebra, with the interactions
given by the simple roots of the algebra. If we denote the fields by
$\phi^i$, and the simple roots of the Kac--Moody algebra by
$\alpha_a^i$, then the action takes the form
\eq
S = \int {\rm d}x^2 \left\{ \hf \partial^\mu\phi\cdd\partial_\mu\phi +
\frac{m^2}{\b^2} \sum_{a} n_a \exp( \beta\alpha_a\cdd\phi )
\right\}
\;,
\label{eq.lag}
\en
where $\b$ is the coupling constant, $m$ is the mass scale and $n_a$
are numbers chosen so that $\phi=0$ is the minimum of the potential.
What makes the Toda theories special is that
classically they are integrable with conserved quantities whose
spins are given by the exponents of the affine algebra. The presence
of higher spin conserved quantities in the quantum theory implies
that the scattering preserves individual particle momenta, and that
the S--matrix factorises on the two particle scatterings.
For details of Kac--Moody algebras and their classification see
e.g. \cite{Kac1}. Each algebra
is denoted $G_n^{(r)}$ where $G_n$ is a finite dimensional Lie
algebra and $r$ is the twist, which can be $1,2$ or $3$.
$G_n^{(r)}$ has $n+1$ simple roots which span $\Bbb R^n$.
To each simple root $\alpha$ of a Kac--Moody algebra we can associate
a dual root,
$\alpha^\vee$, given by $\alpha^\vee = 2 \alpha / |\alpha|^2$. These
dual roots are also the simple roots of a Kac--Moody algebra.
If the algebra and its dual are isomorphic, then we call that algebra
(Langlands) self--dual. The self dual algebras are
$A_n^{(1)},D_n^{(1)},E_n^{(1)}, A_{2n}^{(2)}$, and the non-self dual
algebras come in dual pairs $( B_n^{(1)},A_{2n-1}^{(2)} ),
(C_n^{(1)},D_{2n-1}^{(2)} ), (G_2^{(1)},D_4^{(3)} ), (F_4^{(1)},
E_6^{(2)}) $.

Although all
the divergences can be removed from a two dimensional scalar theory by
normal ordering, at first
sight the action \reff{eq.lag} appears to present difficulties in that
there will be an infinite set of counterterms generated and that these
will not necessarily preserve the form of the action which depends on
only two constants $m,\b$. However, it can be shown that normal
ordering only induces a multiplicative change in the exponentials.
Thus,  for any theory of $n$ fields with $n+1$
exponential interactions all regulation schemes are
equivalent up to
a constant shift in the scalar fields
and a renormalisation of $m$,
 provided that any $n$ of the $n+1$ directions
$\alpha_a$ are linearly independent
\cite{Cole1,DVeg1}. The Affine Toda theories are therefore
renormalisable and it makes sense to discuss the $\b$--dependence of
physical observables since this coupling constant is unaltered by a
change in the regulation scheme.

Since the mass scale $m$ depends on the regulation scheme, it is
only the ratios which are observable. It is a remarkable fact that to
first order in perturbation theory the mass ratios for the self--dual
theories are independent of the coupling constant and keep their
classical values \cite{BCDS1,BCDS4}. This feature enables one to write
down simple S--matrices which have their physical  poles fixed at the
fusion angles given by the classical masses and allowed tree-level  three
point couplings. The S-matrix coupling dependence
is then added in so as to agree with
first order perturbation theory \cite{BCDS1,BCDS4,CMus1}. However
this procedure does not work for
the non-self dual
theories. There are  two reasons. Firstly, the bootstrap constraint on
the S--matrices does
not close on the classical particles; and secondly, the mass ratios
are renormalised by quantum effects even
in first order perturbation theory.
However, by performing a careful analysis of the first order
perturbation theory and requiring  that the two particle S--matrices
obey certain relations (crossing-symmetric, unitarity and a bootstrap
hypothesis),
Delius et al.~were able to postulate the S--matrices of
these theories \cite{DGZa1,DGZa2}. One of the new features of these
S-matrices as compared to the self-dual theories is that  the
position of the poles is now $\beta$ dependent. These S-matrices thus
provide predictions for the mass ratios of the Toda theories in
the large $\b$, non-perturbative regime. The mass
ratios which they found exhibit a relation
between the dual theories: the mass ratios in one theory for
large coupling constant are the same as those in the dual theory for
small coupling constant. Furthermore the S--matrices for two dual
theories may both be expressed as the same function when written in
terms of a parameter $B$ which encodes the coupling constant
behaviour, and the only difference between the two theories is the
dependence of this function $B$ on the coupling constant.
Further evidence for this duality can be found by looking for the
quantum conserved currents \cite{FFre4,DGZa3,KWat3}, which also
exhibit this property; the conserved currents for small $\b$ for one
theory have the same form as the conserved currents for large $\b$ of
the dual theory \cite{KWat3}.

In this paper we present the results of a Monte-Carlo simulation
of the $G^{(1)}_2$ Affine Toda filed theory.
The convention used for the $G^{(1)}_2$ roots was,
\eq
\alpha_1 = (\sqrt{2},0)\,,\;
\alpha_2 = (-1/\sqrt{2},-\sqrt{3/2})\,,\;
\alpha_3 = (-1/\sqrt{2},1/\sqrt{6}) \,.
\en
The corresponding constants $n_a$ are $n_1=2,n_2=1,n_3=3$. The
S--matrix prediction of Delius at al. \cite{DGZa2,Deli1} for the ratio of
the masses of two fundamental particles of this theory is
\eq
\frac{m_2 }{ m_1 } =
\frac{ \sin( 2\pi / H )}{ \sin(\pi / H) }
\,,
\label{eq.func}
\en
where
\eq
H = 6 +  (\b^2 / 2\pi) / ( 1 + \b^2/12\pi )
\,.
\label{eq.H}
\en
This agrees with perturbation theory to one loop. The $G^{(1)}_2$
theory is dual to the $D_4^{(3)}$ theory
 ($H$ flows from 6 to 12 with increasing $\beta$ -- the twist times the
Coxeter numbers of $G_2^{(1)}$ and $D_4^{(3)}$ respectively).
The prediction is that for small $\b$ the mass ratio for the
$G_2^{(1)}$ theory is approximately $\sqrt{3}$, and for large $\b$ it
approaches the classical value of the $D_4^{(3)}$ theory, which is
$\sqrt{ (  \sqrt{3} + 1) / (\sqrt{3} - 1) }$.
We have measured
the flow of the mass ratios as a function of $\beta$. Our observations
strongly support the duality conjecture and are consistent with
the functional form of this flow  \reff{eq.func}.

The rest of the letter is laid out as follows.
In sect.
\ref{secv2} we present the simulation details, and in sect. \ref{secv3}
we summarise our results and compare them with prediction. Finally we
present our conclusions.

\section{Simulation Details \label{secv2} }

A Metropolis simulation was carried out on a periodic square lattice
(most runs using a $64\times 40$ lattice). The discrete Euclidean
action used was
\eq
S= {1\over 2}  \sum_{<nm>} (\phi_n -\phi_m) \cdd (\phi_n
-\phi_m) + \frac{m^2}{\beta^2} \sum_{n,a} n_a
\exp(\beta\alpha_a \cdd \phi_n )\;,
\label{daction} \en
where  $n$ labels a site and the first sum is over nearest-neighbour
sites. Measurements of the zero spatial momentum components of the
three correlation functions
$\vev{\phi^1(x,t) \phi^1(0,0)}$,
$\vev{\phi^1 (x,t) \phi^2(0,0)}$,
$\vev{\phi^2(x,t) \phi^2(0,0)}$ and of the two field
averages $\vev{\phi^1(0,0)}$ and $\vev{\phi^2(0,0)}$ were made.
Labelling the sites by their spatial and time coordinates the
correlation functions were calculated  in the following way (the short
lattice direction is characterised as the spatial direction and the
long length as the time direction):
\eq
\vev{\phi^i (n_t) \phi^j (0)} = {1 \over L} \sum_{m=1}^L ({1\over N}
\sum_{n=1}^N \phi_{n,m}^i )({1\over N} \sum_{n'=1}^N
\phi_{n',m+n_t}^j ) \en
where $N$ and $L$ are the lattice lengths in the spatial and time
directions respectively.
The  matrix of correlation functions was fitted by the function
\eqq
\pmatrix{
\vev{\phi^1 (n_t) \phi^1 (0)} & \vev{\phi^1 (n_t) \phi^2 (0)} \cr
\vev{\phi^1 (n_t) \phi^2 (0)} & \vev{\phi^2 (n_t) \phi^2 (0)}\cr}
=&&\pmatrix{ S_1 S_1 & S_1 S_2 \cr S_2 S_1 & S_2 S_2}
\nonumber\\
+  R(\theta_1) \pmatrix{ a e^{-m_1 n_t}  & 0 \cr 0&0\cr }
R^{(-1)} (\theta_1)
&+&
R(\theta_2) \pmatrix{  0 & 0 \cr 0& b e^{-m_2 n_t} \cr }
R^{(-1)} (\theta_2)
\;, \label{fit}
\enn
where $R(\theta)$ is the rotation matrix
\eq R(\theta) = \pmatrix{
\cos\theta& \sin\theta\cr   -\sin\theta &\cos\theta }
\;,
\en
$m_1$ and $m_2$ the two masses, $a$ and $b$ constants, and  $S_1$ and
$S_2$  correspond to the vacuum expectation values of the fields.
Since we imposed periodic boundary conditions on our lattice, we
should be fitting to a more complicated function that \reff{fit}, with
exponentials replaced by coshes.
In each case, the fitted parameters were
insensitive to this because the statistical errors in the correlation
functions grew with separation.

\blank{
Normally one defines an effective mass for a correlation function
$f(n_t)= \vev{\phi(n_t)\phi(0)}$ as $m_{eff}(n_t) = \log f(n_t+1) -
\log f(n_t)$.
If the correlation function $f(n_t)$ is in fact an exponential then this
effective mass will be constant over the range of $n_t$.
For a system with correlation functions given by \reff{fit}, it is not
possible  to define effective masses
$m_{eff}^1,m_{eff}^2$ from the matrix of correlation functions
which would be constant with $n_t$ and
equal to $m_1,m_2$,  unless $\theta_1 = \theta_2$
}
On fitting to various separations we found that the differences
between the fitted values of $\theta_1,\theta_2$ were of the order of
the statistical errors ($<1$\%). We then chose to fit the correlation
functions with \reff{fit} on separations for which the effective
masses (derived assuming $\theta_1=\theta_2$) were approximately
constant. In practice this meant that fits to \reff{fit} were carried
out over separations $n_t$ from 0 out to some larger distance which
varied between 4 and 15.

The eight parameter fit of \reff{fit}  was carried out with a standard
chi-squared minimisation routine. The chi-squared function involved a
sum over contributions from  the three correlation functions  at
each of the separations being
fitted  plus the two field averages. The errors in the denominator of
the the chi-squared function  were obtained by rebinning the data in
order to allow for computer time autocorrelations (this rebinning
method involves averaging the data into successively fewer bins until
the variance of a  particular parameter over the bins converges
\cite{rebin}). A bootstrap routine was carried out over the binned
data in order to obtain the error estimates on the fitted parameters.
If there were $N_{bin}$ bins of data, then the bootstrap method
involved choosing $N_{bin}$ new bins randomly from these bins (with
repetition when it occurred), fitting our function over this new data
set, and then repeating over an ensemble of such data sets. The
variances  of the fitted parameters over this ensemble gave the quoted
errors.

The simulation was carried out by choosing a particular  $\beta$ value,
and then decreasing $m$ in \reff{daction} until the measured mass ratio
converged to within statistical errors (this is the scaling requirement
that must be met in order that a lattice theory can give information
about the continuum). Finite-size effects were then examined for those
values of $\beta$ for which scaling was observed by
varying the spatial lattice size.

\section{Results \label{secv3}}

The results are collected in figures 1 and 2a -- 2f.
10000 equilibration iterations and 809200 further iterations
were carried out to obtain each mass ratio measurement.
 Each such run on the $64 \times
40$ lattice required approximately
120 hours of CPU time on a Sparc IPC machine.
Figure (1), shows measurements of the mass ratio at two different
masses in equation \reff{eq.lag}, at a selection of $\beta$ values
on the $64 \times 40 $ lattice.
The solid line is the prediction \reff{eq.H} of Delius et al
\cite{DGZa2}.
Figures 2a -- 2f
show all  runs for each of the $\beta$ values.
These latter plots  show the masses plotted against the longer of the
two correlation lengths measured.
In addition, the $\beta=50$ and $\beta=20$ plots (figures 2a and 2b)
show the results of runs carried out
on $64 \times 60 $ and $64 \times 30$ lattices.

In order that scaling is achieved, the points on figures 2a -- 2f
must reach a plateau as the correlation length increases.
The absence of finite-size effects must then be demonstrated
for points that lie on this plateau.

 All the points for $\beta=50$ and $\beta=20$ on the $64 \times 40$
lattice  vary by less than one percent and lie well within each
other's (one sigma) error bars. Furthermore, the results on the
lattices with smaller and larger spatial dimensions agree to the same
accuracy. Thus these two criteria are taken as having been met.

For the other $\beta$ values, the situation is less clear cut. As
$\beta$ was decreased it became necessary to go to larger correlation
length to achieve finite size scaling and the error bars increased in
size due to critical slowing down. Nevertheless for $\beta=5$ and
$\beta=2$ scaling seems to have been achieved. However finite-size
effects may be important at these longer correlation lengths
(finite-size effects were examined in detail for correlation lengths
of approximately 1.5 at $\beta =20$ and 50, compared  to the
correlation lengths of 3-4 relevant here). For $\beta=1$ and
$\beta=0.01$, the perturbative regime, scaling has not been adequately
established (to do this would require larger lattices and longer
runs).

On figure 1 we show a selection of points from the simulations on a
$64 \times 40$ lattice as follows: the shortest two correlation length
points from each of figures 2a and 2b - which lie on the plateaus and
have smaller error bars than the longer correlation length points; the
3 longer correlation points from figure 2c - scaling has not been
reached for the shortest correlation length point; all the points from
figures 2d to 2f - for which  scaling has not been adequately
established.

 From figure 1, the  $\beta=20$ and $\beta=50$ data strongly  support
the duality conjecture (that the mass ration flows to the 1.93 value
shown corresponding to the mass ratio for $D_4^{(3)}$) in the
$\beta\to0$ limit. As discussed above this is the data for which
systematic errors associated with scaling and finite-size effects are
most under control. The rest of the data  is consistent with the
prediction of Delius et al. over the complete $\beta$ value range.



\section{Conclusions}

It is clear that our results are in agreement with the predictions of
duality; that is that the  large $\b$ limit of the mass ratio in the
$G^{(1)}_2$ affine Toda is  the same as small $\beta$ limit of the
mass ratio in  the $D_4^{(3)}$ theory.  In addition  our results for
the intermediate $\b$ range are
not significantly different from the ratio predicted by Delius et al.
\cite{DGZa2}.

Monte-Carlo simulations might also provide a non-perturbative check on
other theoretical predictions. For example
there is some dispute as to the meaning of the anomalous threshold
singularities in the S--matrices of Delius et al. A careful
calculation of the one-loop corrections and of the conservation of the
charges by Delius et al. showed that some of the poles which one would
like to assign to fundamental particles are displaced by loop effects.
It is not clear whether this signals new particles or not.
It should in
principle be able to test whether these do indeed correspond to
physical states by measuring higher-order correlation functions. In
addition it should prove possible to check the proposed form factor
formulae from similar measurements.
Perhaps more immediately, one can investigate whether the observed
duality is present in the $D_4^{(3)}$ theory, and in the other non
self-dual rank two Affine Toda theories. We hope to be able to report
on these questions in the future.

\section{Acknowledgements}

We would like to thank E.~Corrigan for many helpful discussions,
M.~Grisaru for explaining his results, and P.~Craig, A.~Irving and
C.~Michael for advice on the error analysis. RAW thanks the Department
of Applied Mathematics and Theoretical Physics in Cambridge for
their kind hospitality during the period in which the final stages
of this work were completed.
GMTW and RAW acknowledge the support of an SERC research
assistantship and SERC postdoctoral research fellowship respectively.

\begin{thebibliography}{10}

\bibitem{Kac1}
V.~Kac,
\newblock Infinite dimensional Lie algebras,
\newblock Cambridge University Press, 1985.

\bibitem{Cole1}
S.~Coleman,
\newblock Phys. Rev. Lett. D11 (1975) 2088.

\bibitem{DVeg1}
C.~Destri and H.~J. de~Vega,
\newblock Nucl. Phys. B358 (1991) 291.

\bibitem{BCDS1}
H.~W. Braden, E.~F. Corrigan, P.~E. Dorey and R.~Sasaki,
\newblock Phys. Lett. B227 (1989) 411.

\bibitem{BCDS4}
H.~W. Braden, E.~F. Corrigan, P.~E. Dorey and R.~Sasaki,
\newblock Nucl. Phys. B338 (1990) 689 .

\bibitem{CMus1}
P.~Christe and G.~Mussardo,
\newblock Nucl. Phys. B330 (1990) 465.

\bibitem{DGZa1}
G.~W. Delius, M.~T. Grisaru and D.~Zanon,
\newblock Phys. Lett. B277 (1992) 414 .

\bibitem{DGZa2}
G.~W. Delius, M.~T. Grisaru and D.~Zanon{\it,
\newblock Exact S--Matrices for the non-simply-laced affine Toda theories},
\newblock CERN Preprint CERN-TH 6337/91 (1991).

\bibitem{FFre4}
B.~L. Feigin and E.~V. Frenkel{\it,
\newblock Free Field resolutions and affine Toda theory},
\newblock Research Institute in Mathematical Sciences, Kyoto, Preprint
  RIMS--827 (1991).

\bibitem{DGZa3}
G.~W. Delius, M.~T. Grisaru and D.~Zanon{\it,
\newblock Quantum conserved currents in affine Toda theories},
\newblock CERN Preprint CERN-TH 6336/91 (1991).

\bibitem{KWat3}
H.~G. Kausch and G.~M.~T. Watts{\it,
\newblock Duality in Quantum Toda theory and W Algebras},
\newblock Durham University Preprint DTP-92-01 (1992).

\bibitem{Deli1}
G.~W.~Delius, private communication.

\bibitem{rebin}
R.~A. Weston,
\newblock Phys. Lett. B219 (1989) 315 .

\end{thebibliography}

\newpage

\makeatletter
\centering{ Table of collected data}
$$
\begin{array}{|c|c|c|c|c|c|c|}
\hline
\beta & L & m & m_2/m_1 & \delta(m_2/m_1) & m_1 & m_2 \\
\hline
  & 64 \times 60 &   		&1.9208 & 0.0073    & 0.7584 & 1.4568 \\
\@cline[2-2]\@cline[4-7]
  &64 \times 40 &10^{-9}	&1.9302 & 0.0082    & 0.7599 & 1.4669 \\
\@cline[2-2]\@cline[4-7]
50 &64 \times 30 &  		&1.9271 & 0.0077    & 0.7563 & 1.4575 \\
\@cline[2-2]\@cline[3-7]
  &    & 10^{-10}	&1.9317 & 0.0174    & 0.6705 & 1.2952 \\
\@cline[3-7]
  &64 \times 40 & 10^{-15}	&1.9344 & 0.0170    & 0.3629 & 0.7019 \\
\@cline[3-7]
  &    & 10^{-25}	&1.9456 & 0.0155    & 0.3227 & 0.6279 \\
\hline
   & 64 \times 60 &    		&1.9235 & 0.0193    & 0.5873 & 1.1296 \\
\@cline[2-2]\@cline[4-7]
   &64 \times 40 & 10^{-4}	&1.9292 & 0.0108    & 0.5847 & 0.1128 \\
\@cline[2-2]\@cline[4-7]
20 &64 \times 30 &    		&1.9234 & 0.0146    & 0.5846 & 1.1244 \\
\@cline[2-2]\@cline[3-7]
   &    & 10^{-5}   	&1.9330 & 0.0108    & 0.4334 & 0.8377 \\
\@cline[3-7]
   &64 \times 40 & 10^{-6}	&1.9216 & 0.0175    & 0.3217 & 0.8637 \\
\@cline[3-7]
   &    & 10^{-7}	&1.9276 & 0.0224    & 0.2393 & 0.4613 \\
\hline
  &    & 0.1  & 1.8524 & 0.0119    & 0.5156 & 0.8437 \\
\@cline[3-7]
5 &    & 0.07 & 1.8821 & 0.0116    & 0.4294 & 0.8082 \\
\@cline[3-7]
  &64 \times 40 & 0.05 & 1.8832 & 0.0155    & 0.3685 & 0.6940 \\
\@cline[3-7]
  &    & 0.03 & 1.8921 & 0.0180    & 0.2891 & 0.5471 \\
\hline
2 &    & 0.1  & 1.8089 & 0.03      & 0.2203 & 0.3986 \\
\@cline[3-7]
 &64 \times 40 & 0.14 & 1.8029 & 0.0172    & 0.2872 & 0.5179 \\
\hline
  &    & 0.35  & 1.7083 & 0.03      & 0.5255 & 0.8975 \\
\@cline[3-7]
  &    & 0.3   & 1.7151 & 0.0105    & 0.4560 & 0.7822 \\
\@cline[3-7]
1 &64 \times 40 & 0.18  & 1.7365 & 0.0148    & 0.2828 & 0.4911 \\
\@cline[3-7]
  &    & 0.14  & 1.7424 & 0.0294    & 0.2268 & 0.3952 \\
\@cline[3-7]
  &    & 0.12  & 1.7832 & 0.0235    & 0.1889 & 0.3369 \\
\hline
 &    & 0.4   & 1.6898 & 0.017   & 0.5598 & 0.9459 \\
\@cline[3-7]
0.01 &64 \times 40 & 0.25  & 1.7035 & 0.02    & 0.3564 & 0.6071 \\
\@cline[3-7]
 &    & 0.2   & 1.7302 & 0.0260  & 0.2826 & 0.4889 \\
\@cline[3-7]
 &    & 0.15  & 1.7061 & 0.0257  & 0.2149 & 0.3667 \\
\hline
\end{array}
$$

\end{document}

\bye